\documentclass[12pt,preprint]{aastex}

\begin{document}

\shorttitle{}
\title{Deep Optical Imaging of Starbursting ``Transition" Dwarf Galaxies}
\shorttitle{Transition Dwarf Optical Imaging}
\author{Kate E. Dellenbusch\altaffilmark{1}\altaffiltext{1}{Visiting Astronomer, The 0.9m telescope is operated by WIYN Inc.
 on behalf of a Consortium of ten partner Universities and Organizations 
 (see http://www.noao.edu/0.9m/partners.html). WIYN is a joint partnership of the University of Wisconsin 
 at Madison, Indiana University, Yale University, and the National Optical Astronomical Observatory.}}
\affil{Department of Astronomy, University of Wisconsin, 475 N. Charter Street, Madison, WI 53706}
\email{dellenb@astro.wisc.edu}
\author{John S. Gallagher, III}
\affil{Department of Astronomy, University of Wisconsin, 475 N. Charter Street, Madison, WI 53706}
\email{jsg@astro.wisc.edu}
\author{Patricia M. Knezek}
\affil{WIYN Consortium Inc., PO Box 26732, Tucson, AZ 85726}
\email{knezek@noao.edu}
\author{Allison G. Noble\altaffilmark{2}\altaffiltext{2}{Now at McGill University, 3600 rue University, Montreal, QC, Canada; nobleal@physics.mcgill.ca}}
\affil{Department of Astronomy, University of Wisconsin, 475 N. Charter Street, Madison, WI 53706}
\email{agnoble@wisc.edu}

\begin{abstract}

A subgroup of dwarf galaxies have characteristics of a possible evolutionary transition between star-forming systems and dwarf ellipticals.
These systems host significant starbursts in combination with smooth, elliptical outer envelopes and small HI content; they are low on
gas and unlikely to sustain high star formation rates over significant cosmic time spans.  We explore possible origins of such starburst
``transition"
dwarfs using moderately deep optical images.  While galaxy-galaxy interactions could produce these galaxies, no optical evidence
exists for tidal debris or other outer disturbances, and they also lack nearby giant neighbors which could supply recent
perturbations.  Colors of the outer regions indicate that star formation ceased $>$1~Gyr in the past, a longer time span than can be
reasonably associated with the current starbursts.  We consider mechanisms where the starbursts are tied either to interactions with other 
dwarfs or to the state of the interstellar medium, and discuss the possibility of episodic star formation events associated with gas heating 
and cooling in low specific angular momentum galaxies. 

\end{abstract}

\section{Introduction}

The formation and evolution of dwarf galaxies remains poorly understood.  How these processes produce both
rotationally supported (dIrr) as well as dynamically warm (dE/dS0s) dwarf systems remain puzzles,
despite considerable effort.  An outstanding question in dwarf galaxy evolution is whether there is an evolutionary
connection between the various morphological classes of dwarf galaxies (e.g. van den Bergh 1977, van Zee et al. 1998,
van Zee et al. 2001, Lisker et al. 2007).  Specifically,
can dwarfs of one morphological type evolve by some process into a different type?  For example, do some dIrrs
evolve into dE/dS0s or did the dE/dS0s we see in the Universe originally form as that type?  These questions
regarding the formation and evolution of dwarf galaxies are of particular importance in light of hierarchical cold
dark matter (CDM) galaxy formation models which predict large numbers of dwarf galaxies and suggests that larger
galaxies are built up from the accretion of many low mass halos (e.g. White \& Frenk, 1991).

A related issue concerns the impact of star formation on dwarf galaxy evolution and particularly the role of
episodic star formation (e.g. Searle, Sargent \& Bagnuollo 1973, Lee et al. 2002).  How does star formation relate to
morphology in low mass galaxies?  For example, where do blue compact
dwarfs (BCDs) fit in the dwarf galaxy zoo as starbursting objects and are they related to either dIrrs or dE/dS0s
(e.g. Sung et al. 1998, Gil de Paz \& Madore 2005, Noeske et al. 2005)?  In particular, it has been suggested that 
BCD galaxies may evolve into dE/dS0 galaxies after they lose their gas, either through supernovae-driven winds
during episodes of intense star formation (Marlowe et al. 1999, Tajiri \& Kamaya 2002), or through stripping processes induced by galaxy-galaxy interactions
(Drinkwater \& Hardy 1991).  Or, are some of these objects dEs where star formation has been renewed through gas capture
as suggested by Silk, Wyse, \& Shields (1987)?  Previous studies, which have concentrated mainly on optical
morphological differences between dwarf galaxies and to some extent on their neutral gas content, have not firmly 
established a direct evolutionary connection between dIs and dEs in galaxy groups and other low density environments (e.g. van Zee, 
Salzer \& Skillman 2001, Noeske et al. 2005).  

Even in the Local Group the situation is unclear.  Many Local Group dwarfs have complex star formation histories
and kinematic peculiarities, and at least five of the faint dwarfs possess optical morphologies which place them in a 
``transition" or ``mixed-type" structural category between dI and dE/dS0 (e.g. Mateo 1998, Grebel, Gallagher, \& Harbeck 2003).
While nearby dE galaxies, such as NGC~185 and NGC~205 in the local group, support star formation, the rates are
extremely low and consistent with these objects' small gas supplies.

More distant galaxy samples also contain dwarfs with early-type structures that also host HI gas and often also some young stars.
As in the Local Group, these transition dwarfs frequently are small, low luminosity objects containing $\lesssim$ 10$^6$ M$_{\odot}$
of HI and having correspondingly low star formation rates (e.g. Bouchard et al. 2005).  On the other hand, the Virgo Cluster of galaxies
appears to contain a complete sequence of more luminous dwarfs with declining and increasingly concentrated star formation, extending from typical
BCDs through to blue core dEs (Gallagher \& Hunter 1989, Lisker et al. 2006,2007).  While moderate luminosity galaxies with early-type
outer structures and young stellar populations in their centers also exist in less dense environments (e.g. NGC~404, del Rio et al. 2004;
NGC~5102; Deharveng et al. 1997), they are sufficiently rare that it is difficult to determine if evolutionary sequences exist.

In this paper we explore possible histories for a sample of actively star forming luminous dwarf galaxies with transitional 
properties between BCDs and dEs residing in loose
group environments.  These systems are fairly isolated and show little indication of a recent interaction or merger.  
They have an intriguing combination of characteristics which do not allow them to 
easily be categorized as either dI/BCD or dE systems, but rather show a mixture of the two.  The key properties
which indicate these low luminosity galaxies {\it may} be in the midst of an evolutionary transition from dI/BCD to dE
include: 
\begin{itemize}
\item They are actively forming stars with star formation rates between 0.1 and 1 M$_{\odot}$/yr.
\item The star formation is currently centrally concentrated, with the outer regions composed
of an older stellar population.  
\item Although actively forming stars, the starburst is fueled by very little HI
gas, with M$_{HI}/L_B \lesssim$\ 0.1.  This compares with M$_{HI}/L_B >$ 0.2 for typical dIs (Roberts \& Haynes 1994) and BCDs (Pustilnik et al. 2002) and 
M$_{HI}/L_B <$ 0.1 for dEs.  Star formation can continue in our sample galaxies only for about another 10$^9$ years at their current rate,
based on their HI content; in a few cases, including molecular gas will at most double this 
time (e.g.  Jackson et al. 1989, Gordon 1991). 
\item Unlike many other star forming dwarf galaxies including BCDs they have high oxygen
abundance ratios, with 12 + log(O/H) $>$ 8.4 as we found in Dellenbusch et al. (2007).
\item Their outer optical colors are similar to those of typical BCDs with (B-R)$_0$ of order 1 (Gil de Paz et al. 2005), but bluer than
the (B-R)$_0$ $\approx$ 1.3 - 1.4 expected for slightly fainter dEs (Conselice et al. 2003).  In this regard our sample resembles the Gil
de Paz et al. "E-type" BCDs.  
\item  These objects have
smooth outer isophotes which are much more indicative of early- rather than late-type galaxy structures.
\end{itemize}
Table~\ref{properties} quantifies several of these properties.

In this paper we examine the significance of these final two characteristics.  In \S 2 we discuss
the observations and data reduction.  In \S 3 we describe our analysis and modeling processes 
used to examine the data for evidence of tidal debris and fine structures.  In this section we also describe
the results of this analysis.  We discuss two possible evolutionary pictures to explain the galaxies' star 
formation histories as a means of transition in \S 4.  \S 5 contains a brief summary and our conclusions.

\section{Observations and Data Reduction}

For this study, we obtained imaging data with the WIYN 0.9-m telescope on Kitt Peak in Arizona.  
We used the S2KB CCD camera to obtain deep R-band imaging of five transition dwarf galaxies.  The
S2KB camera provides a 20' $\times$ 20' field of view with a plate scale of 0\farcs6.  The data 
were taken on 24-26 and 28-29 April 2006; conditions were not photometric.  We obtained total exposure times of 50 to 62 minutes for
each galaxy by combining 4 or 5 750 second exposures.  The data were reduced using standard methods in 
IRAF\footnote{IRAF is distributed by the National Optical Astronomy
Observatories, which are operated by the Association of Universities for Research in Astronomy, Inc., under 
cooperative agreement with the National Science Foundation.}.  Data reduction included subtraction of combined
bias frames and flat fielding using combined dome flats.  The multiple dithered frames were aligned and combined.

We also use short B- and R-band imaging observations made at Michigan-Dartmouth-MIT (MDM) Observatory on the 1.3 m McGraw-Hill telescope
during 10-12 April 1996.  Typical integration times were 20-39 minutes in B, and 10-15 minutes in R under generally photometric conditions.  Details of the data reduction
performed on these images can be found in Knezek, Sembach, \& Gallagher, 1999).

The sample of transition galaxies studied here, consisting of NGC~3265, NGC~3353, NGC~3773, NGC~3928, and IC~745, was
selected to have a mixture of early- and late-type dwarf galaxy characteristics.  That is, an underlying dE or dS0 with a high 
rate of current star formation.  They thus were chosen in the spirit of Sandage \& Hoffman (1991) and were initially observed
as transition dwarf candidates with MDM along with NGC~3377A, NGC~4286, and IC~3475, which were discussed in Knezek, Sembach, \&
Gallagher (1999).

\section{Results and analysis}

\subsection{Elliptical Modeling and Residual images}

In order to study the morphology and fine structure of our galaxies, we produced residual images, in other
words the difference of the original images minus galaxy models.  This allows weak features which are otherwise
dominated by the overall smooth and symmetric light distribution to be more easily discernable.

These residual images are created using several routines in the IRAF package ISOPHOTE.  First, elliptical 
isophotes are fit to the galaxies using the ELLIPSE routine.  The results from such a model of a galaxy's 
surface brightness distribution is then input into the BMODEL task to create a two-dimensional model 
of the galaxy (e.g. Forebes \& Thomson 1992, Lisker et al. 2006), representing the smooth elliptical structure of each system.
Because of intense star formation in the very centers of these galaxies, ELLIPSE does a poor job of fitting isophotes
at small radii.  Ellipses were fit starting in the outer galaxy and proceeding inwards.  These fits were stopped inside the 5 arcsecond
isophote.
By allowing the ellipticity and position angle of the model-fit to vary, any large-scale structure such as
a bar, are included in the model.  When the model is then subtracted from the original galaxy image,
the residual more clearly reveals any fine structure (e.g. dust, tidal tails, shells) which had been 
hidden beneath the dominant elliptical light distribution.
 
\subsection{Dwarf Galaxy Fine Structure Features}

\begin{figure}
\plotone{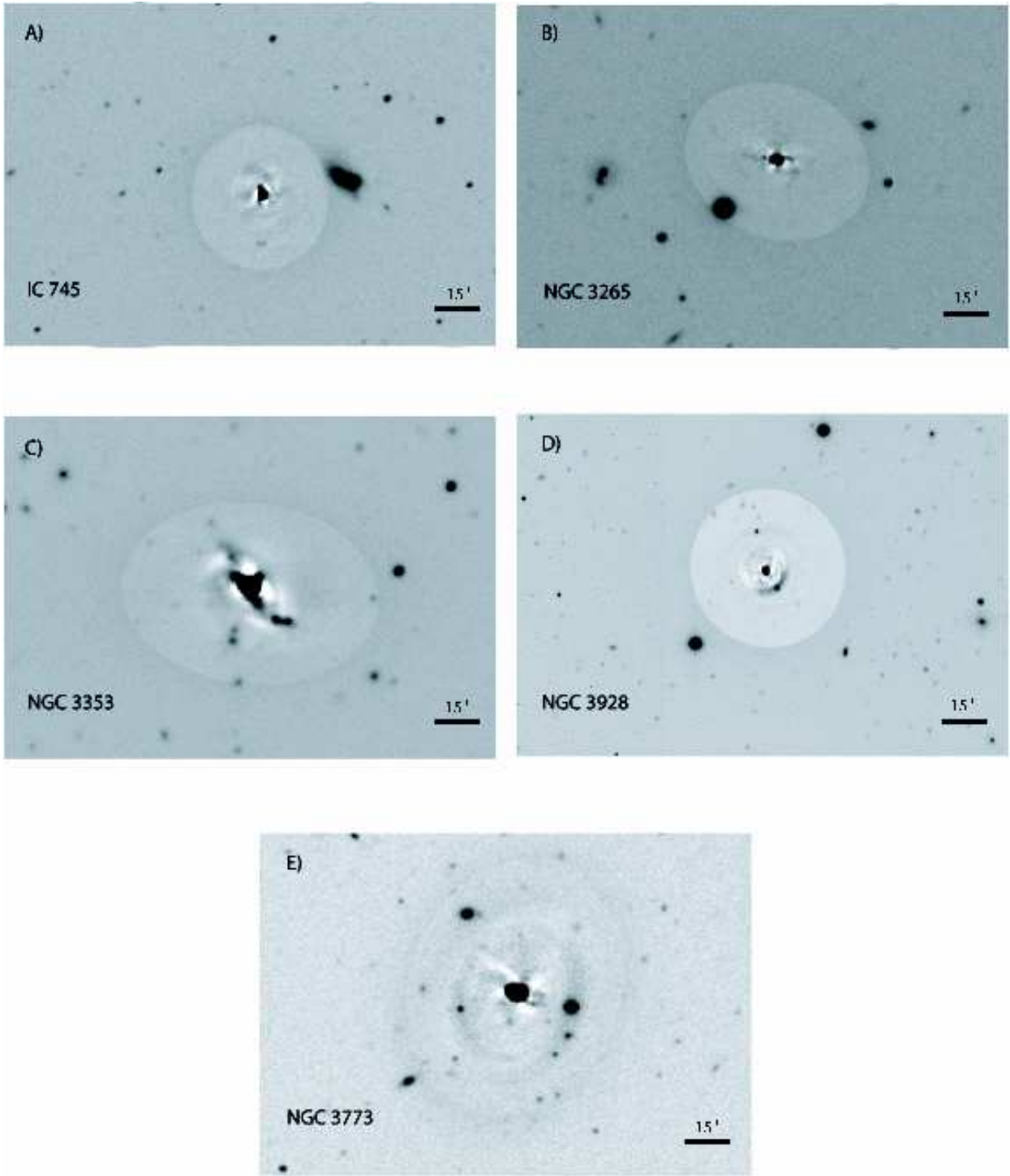}
\caption{Residual negative images in R-band of galaxies: A) IC~745, B) NGC~3265, C) NGC~3353, D) NGC~3928, and E) NGC~3773.  Dust features appear as lighter
regions in these negative images.  Because of the intense star formation there, the inner $\sim$ 5 arcseconds of each galaxy were not included in 
the model and therefore appear as dark regions. \label{bmodel}}
\end{figure}

\begin{figure}
\plotone{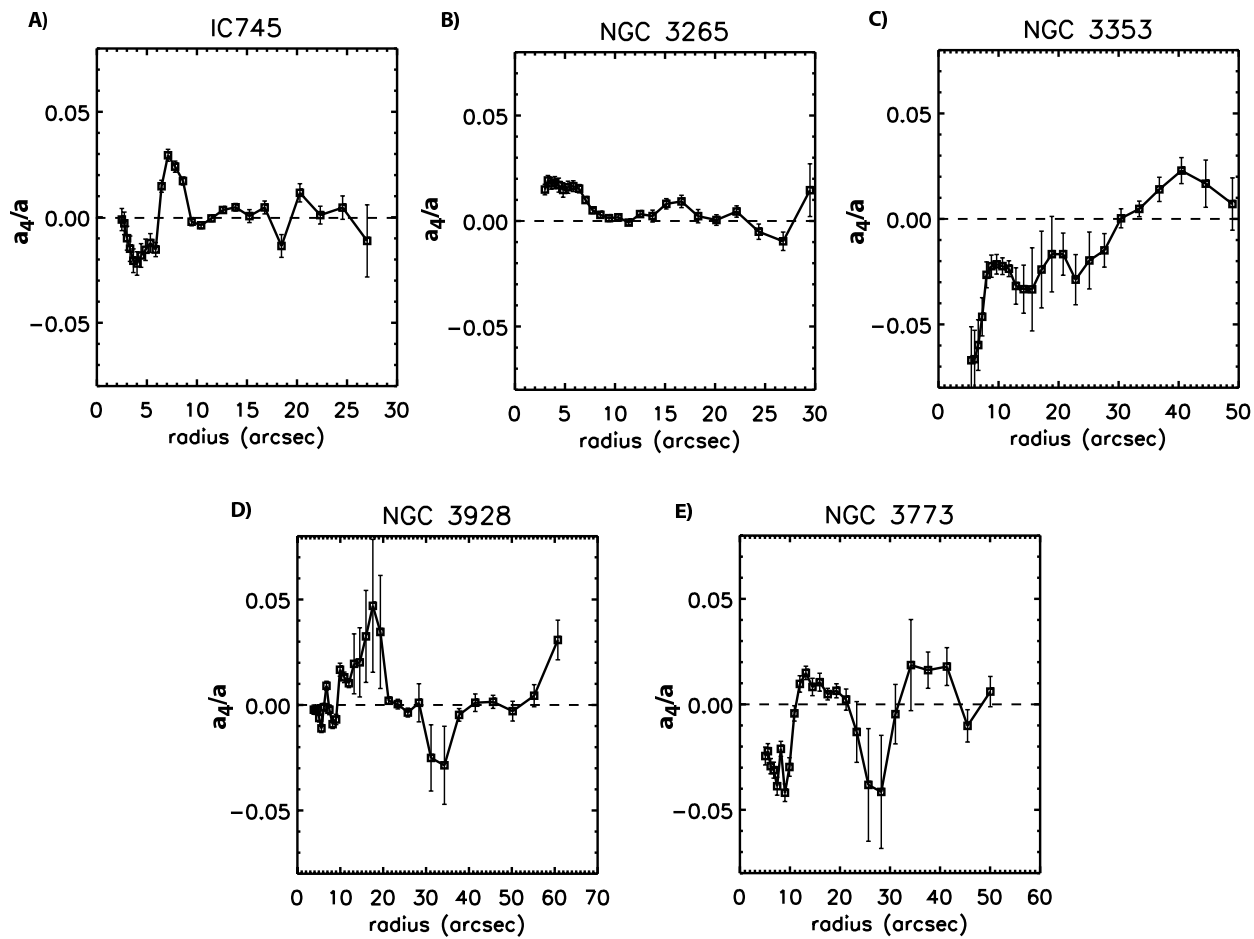}
\caption{Deviation from perfect ellipticity, described by the Fourier a$_4$ parameter, as a function of distance along the semi-major axis 
for galaxies A) IC~745, B) NGC~3265, C) NGC~3353, D) NGC~3928, and E) NGC~3773.  a$_4$ $>$ 0 describe disky isophotal shapes while a$_4$ $<$ 0
describe boxy isophotes.  Although our images are moderately deep, S/N becomes low at outer isophotes and therefore we terminate the plot
at $\sim$ 30 arcseconds for the two smaller galaxies.  \label{A4}}
\end{figure}

The procedure described above was used to examine the deep R-band images for evidence of tidal interaction and clues to their evolutionary
histories. We do not find tidal tails or extended tidal debris in any of the galaxies studied.  The residual images do however reveal
interesting fine structures near the galaxy centers where we find the light distribution is not 
smooth. Several of them have clear dust lanes, in particular NGC~3773 and IC~745 (See 
Figure~\ref{bmodel}).
The residual images in Figure~\ref{bmodel} are presented as negative images.  The dust features therefore appear as lighter regions.
These dust features are also visible in the original images.
In addition, most of the objects do not show significantly boxy or disky isophotes as derived from a$_4$ coefficients of 
the ellipse fits.  NGC~3353 does show evidence of being boxy with -0.07 $\lesssim$ a$_4$/a $\lesssim$ 0
out to a radius of 30 arcsec, while NGC~3265 is weakly disky with 0 $\lesssim$ a$_4$/a $\lesssim$ 0.02 inside of 10 arcseconds.  
IC~745, NGC~3928, and NGC~3773 are not predominantly boxy
or disky  with a$_4$/a not trending toward positive or negative values on average (See Figure~\ref{A4}).  Thus, the conservative result is that the outer isophotes 
are quite close to ellipses.  It is important to note however, that these trends are likely influenced by the prominent dust features
present in the inner parts of these objects.  In addition to dust features, several of the galaxies have complex fine structures in their 
inner regions (See Figure~\ref{bmodel}), indicating the light distribution is not smooth in the inner parts of these galaxies.  

Galaxies that have not experienced mergers in the past several Gyr generally have symmetric stellar bodies.  In systems where mergers or
stellar mass 
transfer events are more recent, distinctive features can be seen, including tidal tails.  As these systems age, structural features associated with past
events become increasingly subtle, passing from stellar shells and ripples to possibly tidal bars (Miwa \& Noguchi 1998) or boxy isophotes 
(see Barnes \& Hernquist 1992).  {\it If} the galaxies we observe as starbursting transition dwarfs formed via near equal mass mergers, then these events
were in the distant past or involved a companion with very few stars.

The fine structure features as described above, are apparently associated with gas and are similar to those others have found in
larger isolated elliptical galaxies.  For example, Reda et al. (2004) studied a sample of nearby isolated early-type galaxies and found a variety of fine
structure features in several of them such as dust lanes and shells as well as both disky and boxy isophotes.  For several of the E's studied, Reda et al.
suggest they are the remnant of a galaxy merger, as their light distributions deviate from perfect ellipticity.  Thus the structure of the ISM in our sample
allows for the possibility of past galaxy mergers; however if these events occurred, they took place many outer-galaxy dynamical time scales in the past.

\section{Discussion}

The galaxies discussed in this paper exhibit a puzzling combination of properties.  Although these types of galaxies are not 
ubiquitous in the local universe, they may represent an important snapshot in dwarf galaxy evolution.  It is therefore
useful to understand their possible evolutionary histories and likely future states.  Two potential scenarios are
explored here; an internal ISM instability leading to multiple starburst episodes and a past interaction or dwarf-dwarf
merger event.

\subsection{Internal Chemo-Dynamical Star Formation Regulation}

The pattern of star formation in these galaxies indicates that producing a concentration of low angular momentum gas in the central regions plays an
important role in their evolution. In this 
regard the problem is similar to that generally seen in BCD galaxies with high star formation rates: how is the central region 
of a starbursting dwarf fueled with gas (e.g. Taylor et al.\ 1993, van Zee et al.\ 2001)? 

In galaxies with low specific angular momenta, random motions and thermal energy could play important roles in supporting a spatially extended ISM.  
In the extreme situation where these are the only factors, numerical simulations suggest that the system can become unstable against bursts of star formation.
This idea was first explored by Loose et al.\  (1982) who modeled the Galactic center. Stars energize the ISM which then expands, shutting off star formation,
leading to collapse and another round of star formation; a cycle that resembles a kind of cosmic diesel engine.  This type of galactic ``star formation 
dieseling" also is seen in some of the more sophisticated 1-dimensional simulations of non-rotating dE galaxies by Hensler et al.\ (2004; see also Pelupessy et al. 2004). 
The basic principle
is again the ability of mechanical energy from stars to expand the radial extent of a low mass ISM to the point where star formation declines, leading to a 
collapse on a gas dissipational time scale and another round of star formation, or gas exhaustion      

It is not clear whether the kinds of star formation rate oscillations found in these models occur in galaxies.  A connection between 
mechanical power and the overall radial structure of the ISM is one key feature.  For example, Tajiri \& Kamaya (2002) estimate conditions under which 
BCD HI envelopes could be blown away, and the galaxies in this sample sit near their boundary SFR for retaining their HI. Thus we expect the envelopes may 
be influenced by the current starbursts. A full assessment, however, requires knowledge of the gravitational potentials and structural form of the system.
Unfortunately the stellar kinematics of the galaxies in our sample are not 
known and thus more detailed modeling is not yet warranted.  

Studies of HI in galaxies like those in our sample (central starburst, moderate luminosity, low $M_{HI}/L$), however, show that some of the gas is in 
a rotating disk surrounding the central gas concentration (Lake et al. 1987, Sadler et al.\ 2000, van Zee et al.\ 2001). In some cases the gas kinematics are 
not regular; e.g. Lake et al. note the HI in the outer parts of NGC~3265 has confused kinematics, which they summarize as ``infall, outflow, no-flow''.  
Lake et al.\ and Sadler et al.
\ consider that some of the HI could have been recently accreted, but this now seems unlikely given the high gas phase chemical abundances that we find in 
Dellenbusch et al.\ (2007). 

The possibility of an internal instability that leads to an episodic central concentration of gas is consistent with the chemical
abundances and therefore merits closer examination. 

An alternative internal model for feeding gas into the centers of low mass galaxies via viscosity of 
gas in their disks has been developed by Noguchi (2001). In this picture low 
density gas disks in small galaxies fail to efficiently form stars, thereby allowing gas to accrete into the centers of the systems. This model, however, 
produces long-lived central star forming zones and thus it is not clear if it can be applied to the low HI content, high SFR objects studied 
here.

\subsection{Old Interaction/Merger Event}

An alternative mechanism to explain the evolution of these galaxies is a past interaction with another galaxy or possibly a 
dwarf-dwarf merger.  

The amorphous dwarf galaxy NGC 5253 is a well studied nearby starburst galaxy which is similar in several ways to the galaxies
discussed in this paper. It has been suggested that the central starburst in NGC 5253 may have been triggered by an interaction with M83 
(e.g. van den Bergh \ 1980).  NGC 5253 is an amorphous dwarf galaxy with an underlying elliptical structure and bright HII regions 
in the core.  Like the transition dwarfs in this study, it has a low neutral gas mass-to-light ratio of M$_{HI}/L_B$ = 0.04 and an H$\alpha$ 
derived star formation rate of 0.12 M$_\sun$/yr (Ott et al. \ 2005).  It does however have a somewhat lower oxygen abundance 
with 12+log(O/H) = 8.23 (Martin \ 1997).  

Kobulnicky \& Skillman (1995) have suggested the current star formation activity of NGC 5253 may be due to the accretion of 
a gas-rich companion or the result of an interaction with nearby M83 about 10$^9$ years ago.  M83 and NGC 5253 are separated
by only 0.15 Mpc thus making an interaction between them $\sim$ 1 Gyr ago possible.  This is supported by the galaxy's unusual gas dynamics
(Kobulnicky \& Skillman 1995) and an elliptical young halo stellar population  with ages of 10$^8$ - 10$^9$ yrs (Caldwell \& Phillips 1989).

Could similar interactions be responsible for triggering the star formation we see in our transition dwarf sample?   First, are there
any galaxies nearby with which the transition dwarfs could have interacted?  To examine this we must first estimate when stars last formed 
in the outer regions of these galaxies.  We currently only have one color, (B-R), from short exposures obtained with the 1.3 m McGraw-Hill
telescope at Michigan-Dartmouth-MIT Observatory in addition to the deep R-band data presented here.  Although more colors would be beneficial,
we can roughly estimate the minimum time since stars formed in the red outer regions of the galaxies using these data and spectral evolutionary synthesis models.

An estimate of the minimum time passed since the population of stars in the outer disk formed was made using models from the
G\"ottingen Galaxy Evolution (GALEV) Group courtesy of Dr. Uta Fritze-v. Alvensleben.  These are spectral evolutionary synthesis models for 
several metallicities and include gaseous emission (Schulz et al. 2002, Anders \& Fritze-v. Alvensleben 2003).  We use data from these
models to plot the variation of (B-R) color with time for a galaxy which has experienced a burst of star formation.  An estimate of the time
since stars last formed in the outer galaxy envelope can then be made based on the time elapsed since the model burst for the color of the simulated galaxy to fade to the observed outer region color.  
This was done using
GALEV models with a metallicity of z=0.008 for the underlying galaxy and a burst of solar metallicity (z=0.02).  Constant and exponential
star formation rates were explored for the underlying galaxy model.  In both cases a burst with a simple stellar 
population was placed on top of the underlying galaxy population.  In addition, models for different burst strengths were also examined.

\begin{figure}
\plotone{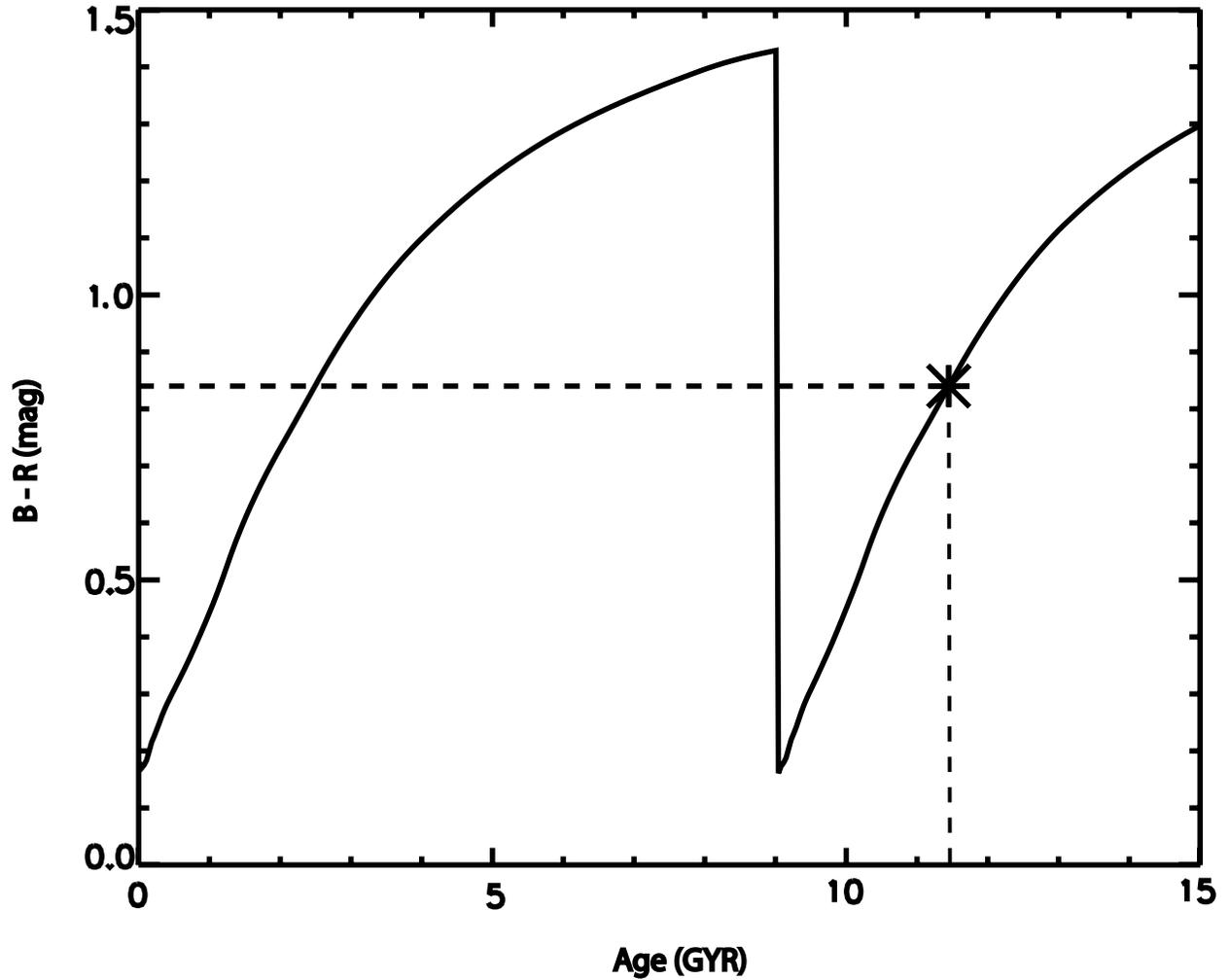}
\caption{Illustrative spectral evolution synthesis model including an underlying elliptical galaxy model having an exponential star formation
rate with a simple stellar population starburst occurring after 9 Gyr (Solid Curve).  Horizontal dashed line indicates B-R color
of the quiescent outer envelope of NGC~3353.  The model indicates that for the color to have faded to B-R=0.84, after an instantaneous burst of
star formation, 2.5 Gyrs have passed (Vertical Dashed Line).  \label{burstage}}
\end{figure}

We find the model post-burst colors to be too blue for the case of constant star formation rate.  Observed integrated (B-R)$_0$ colors
for our galaxies range from 0.84 to 1.2 (see Table~\ref{properties}), while the model post-burst B-R color only gets as red as B-R=0.7 by 6 Gyr after
the burst.  Modeling the underlying galaxy with an exponential star formation rate of the form ($SFR \propto\  e^{-t/1Gyr}$) produces
post-burst colors which are a much better fit to our outer colors data.  Changing the strength  of the burst however, does not significantly affect our age 
estimate.  As a result of these models, we find that at least between 2 and 6 Gyr have passed since star formation ceased in the outer regions
of the galaxies studied here.  An example of one of our models is shown in Figure~\ref{burstage} with (B-R) indicated for the envelope of 
NGC 3353.  These results are consistent with ages found for the underlying stellar population of NGC 3353 by Cair{\'o}s et al. (2007). 

Although these galaxies have no obvious close companions with which they are clearly interacting, they are not completely 
isolated.  The galaxy which visually appears near IC~745 in Figure~\ref{bmodel} is actually a background object.  
Assuming a relative velocity of 200 km/s between galaxies in the loose group environments in which the transition dwarfs
reside, in the estimated time elapsed since the outer regions last experienced star formation, they could now be separated by about
$\sim$0.5~Mpc.  Several other galaxies can be found within 1 Mpc of each of our sample galaxies, and smaller 
gas-rich companions may be relatively common but difficult to locate in optical data (Taylor et al. 1995, 1996, Hogg et al. 1998, Noeske et al. 2001, Pustilnik et al. 2001).  

Any proposed mechanism for producing these dwarfs must be able to explain the observed low gas content and centrally concentrated star formation as well as envelope structure 
and colors.  A likely scenario
therefore is a mechanism through which some gas can be removed from the galaxy and much of the remaining gas falls into the center.  An interaction could
be a way to achieve this.  For example, Bushouse (1987) found that the majority of interaction induced star formation is concentrated near
the nuclei of disk galaxies and that gas depletion times are lower than for isolated spirals.

For the gas to be moved toward the center, it must lose angular momentum and dissipate energy.
Numerical simulations of spiral arms and bars tidally induced in disk galaxies (e.g. Noguchi \& Ishibashi 1986, Noguchi 1987) and 
Noguchi (1988) showed that the gas is also affected, whereby the stellar bar induces the infall of gas toward the galactic nucleus.  However, most
of the transition dwarfs examined here do not appear to have such features, although it is uncertain what the bar lifetime might be for such objects.
An exception to this is NGC 3353, which has star formation
occurring in an elongated arc, as seen in H$\alpha$ imaging (Dellenbusch et al. 2007 in prep.), that is not as centrally concentrated as in the 
other sample galaxies.  This may be a bar-like structure.  In 
addition, van Zee et al. (2001) found the HI in many BCDs, which have characteristics in common with transition dwarfs,
to have intrinsically low angular momenta.

Another related scenario to explain the evolution of transition galaxies is a dwarf-dwarf merger.  Although dwarf galaxy mergers with larger systems
have been modeled (e.g. Hernquist 1989, Mihos \& Hernquist 1994), the results of which can also describe a dwarf galaxy accreting a cloud for example,
there is little discussion of dwarf-dwarf merger simulations in the literature.  This is no doubt at least in part because they occur much less frequently than
other types of merger events.  However, there is evidence that they take place.  One example is II~Zw~40, a BCD which is thought to be an on-going
merger of two gas-rich dwarf galaxies.  Optically, II~Zw~40 is dominated by a starburst with two faint tails extending outward
(Sargent \& Searle 1970, Baldwin, Spinrad, \& Terlevich 1982).  It's HI extends several times the optical diameter with distinct tidal tails extending even
further.  As van Zee, Skillman, \& Salzer (1998) note, the morphology of the system, a strong color gradient indicating the low surface brightness
optical tidal tails are from an older stellar population, and no evidence for a second optical galaxy along the HI tidal tails indicate II~Zw~40
is likely a merger rather than just a tidal interaction.  

Tidal interactions have often been invoked as a mechanism by which star formation can be triggered in a galaxy as well as an explanation for the high 
star formation rates found in BCD's (e.g Brinks \& Klein 1988).  Another possible dwarf-dwarf merger is the galaxy He~2-10 which is a Wolf-Rayet 
galaxy with a primarily elliptical appearance and two bright star clusters.  It also exhibits a tail-like feature in both HI and CO.  
Kobulnicky et al. (1995) suggest, based on HI and CO kinematics, that He~2-10 is an advanced merger of two dwarf galaxies.  A third well known
example of two dwarfs interacting are the Small and Large Magellanic Clouds.  Could the transition dwarfs which we are studying be a more evolved
version of a dwarf-dwarf merger where it is no longer morphologically apparent a merger has taken place?

Although dwarf-dwarf mergers occur in nature and may provide an explanation for at least some transition dwarfs, they are rare.  This
is to be expected because dwarfs have small interaction cross-sections.  Also, for a merger to actually take place, the two
interactors must have very low relative velocities.  As galaxies with transition dwarf characteristics also seem to be rare, it is possible
we are seeing an uncommon form of dwarf-dwarf merger in isolated environments with energy dissipation playing an important role in the
subsequent evolution of these systems.

Although the transition galaxies discussed here do not appear very disturbed and have regular outer isophotes, we cannot rule out an interaction or
merger as playing an important role in their evolutionary and star formation histories.  Several dynamical crossing times have passed since star
formation occurred in the outer envelope, and it is therefore reasonable for minimal evidence of tidal interaction to remain today.  
On the other hand, the very long time scales that we find seem to be at odds with the intrinsically short time scales expected for starburst development 
in small galaxies (e.g. Recchi et al. 2002).

\section{Summary \& Conclusions}

In summary, we have presented results of deep R-band imaging for five starbursting transition dwarf galaxies.  Our observational results show:

\begin{enumerate}
\item All five galaxies exhibit smooth elliptical outer isophotes.
\item We find no evidence of extended tidal debris or other indications of a recent major interaction (e.g. tails, shells or ripples).
\item Fine structure exists in the central regions that are associated with the ISM and star formation, with dust structures and HII 
regions being prominent in several of the galaxies.
\item Outer envelope colors are consistent with having no star formation in the outer regions for the past several Gyr.  This timescale
could be longer if star formation slowly declined rather than stopping suddenly.
\end{enumerate}

We consider two possible classes of evolutionary histories which could be consistent with the observed properties and locations of 
transition dwarf galaxies:  an internal ISM instability and a past interaction or merger event.  The interaction or dwarf-dwarf merger
scenario remains a possibility, although we find no strong evidence to support it.  After several billion years it is not clear if
we should still find evidence of tidal debris in the outer regions of the galaxies.  However, long term products of dwarf interactions 
and dwarf-dwarf mergers in particular have not been well modeled (e.g. the effects of large dark matter contents on merger products). 

The other possibility of an unstable ISM seems promising to explain the central star formation in these objects.  In this case the eventual
evolution to an inactive state will be slower, possibly requiring several starburst cycles.  Observationally the model can be tested
by looking for the young stellar populations in fading starburst phases in dE-like dwarfs.

A remaining question, which is particularly puzzling, is the issue of timescales.  If these objects last experienced significant star
formation in their outer regions $\gtrsim$ 1 Gyr ago, why are they currently undergoing bursts of centrally concentrated star formation?  
What mechanisms control the structure of the ISM such that these galaxies now support rapid star formation, and will these circumstances
lead to galaxies with little or no star formation, i.e. objects resembling dEs, within the next $\sim$ 1 Gyr?

\acknowledgments

We would like to thank P. Mucciarelli for assisting with the observations. In addition we wish to thank H. Schweiker and J. Davies for their
assistance and dedication to the WIYN 0.9m telescope.  We also thank U. Fritze-von Alvensleben for the use of the GALEV models and helpful
discussions regarding them.  This research is supported in part by the University of Wisconsin-Madison Graduate School.

\clearpage

\begin{deluxetable}{lcccccc}
\tablewidth{0pt}
\tablecaption{Galaxy Properties \label{properties}}
\tablehead{
\colhead{Galaxy} &
\colhead{D(Mpc)\tablenotemark{\dag}} &
\colhead{SFR (M$_{\odot}$yr$^{-1}$)} &
\colhead{L$_B$ ( x 10$^8$L$_{\odot}$)} &
\colhead{M$_{HI}$/L$_B$} &
\colhead{12 + log(O/H)} &
\colhead{(B-R)$_0$\tablenotemark{\ddag}}
}
\startdata
IC 745 & 15.3 & 0.2 & 4.9 & 0.05\tablenotemark{a} & 9.0 & 0.92 \\
NGC 3265 & 17.6 & 0.1 & 7.8 & 0.10\tablenotemark{b} & 9.2 & 1.19 \\
NGC 3353 & 12.6 & 0.8 & 7.8 & 0.30\tablenotemark{c} & 8.4 & 0.84 \\
NGC 3928 & 13.2 & 0.1 & 3.6 & 0.33\tablenotemark{d} & 9.1 & 1.19 \\
NGC 3773 & 13.2 & 0.1 & 6.5 & 0.08\tablenotemark{e} & 8.8 & 0.88 \\
\enddata

\tablenotetext{\dag}{ A value of H$_0$ = 75 km/s/Mpc is used to calculate distances}
\tablenotetext{\ddag}{ Outer region color; integrated colors will be bluer}

\tablenotetext{a}{Martin 1998}
\tablenotetext{b}{{L}ake \& Schommer 1984}
\tablenotetext{c}{Gordon \& Gottesman 1981}
\tablenotetext{d}{{L}i et al. 1994}
\tablenotetext{e}{Burstein, Krumm, \& Salpeter 1987}

\end{deluxetable}

\clearpage

\clearpage

\clearpage

\end{document}